          [1999/12/02 1.01 (PWD)]
\documentclass[a4paper,twocolumn,]{esapub} % European paper
\input{epsf}
\usepackage{graphicx}
\usepackage{natbib}

\title{Quark stars in Low-Mass X-ray Binaries: for and against
}
\author {W. Klu\'zniak}
\author {T. Bulik}
\author {D. Gondek-Rosi\'nska}
\affil{Copernicus Astronomical Center, Bartycka 18, 00-716 Warszawa, Poland }

\begin{document}


\maketitle

\begin{abstract}
  It has been suggested that both X-ray bursters and millisecond
radio pulsars may be strange (quark) stars, rather than neutron stars.
Confirming (or rejecting) this suggestion may require knowing what
role strong-field effects of general relativity play in the accretion flow
of the compact X-ray source in low-mass X-ray binaries
(LMXBs).
We discuss the range of rotational and orbital frequencies,
and of masses expected in various models of strange stars, and  compare
them with observational constraints, suggested by
the observed frequencies of kHz QPOs. We explain why future
observations of transients (such as SAX J~1808.4-3658) may be crucial
to understanding the precise nature of the accreting source.
For flattened (e.g., rapidly rotating) distributions of matter,
an innermost (marginally) stable orbit may be present even if relativistic
effects are negligible. Depending on the stellar rotation rate,
the same value of orbital frequency in the innermost stable orbit
(say, 0.9 kHz) can correspond to a star of mass equal to $2.4M_\odot$,
$1.4M_\odot$, $0.1M_\odot$, $0.01M_\odot$, or less.

\end{abstract}

\section{Introduction}
Bodmer (1971) suggested that quark matter (composed of roughly
equal number of up, down, and strange quarks) is the ground state of
matter. Brecher and Caporaso (1976), Witten (1984),
Alcock et al. (1986), Haensel et al. (1986), and others, have
computed the structure of gravitating objects composed of such
matter, the so called strange stars.
  Because the density of quark matter would exceed only slightly
nuclear density, the radii of these quark stars would be comparable
to the radii of neutron stars, as would be the maximum masses---it would
be hard to distinguish the two at first sight.
Although young glitching radiopulsars are likely to be neutron stars
(Alpar 1987), we are unaware of any reason why either the observed
millisecond pulsars, or X-ray bursters (as well as the ``Z sources'') in LMXBs,
should not be quark stars instead.

Unlike for neutron stars, which can exist only above a certain mass
($\sim 0.1M_\odot$), there is no lower limit to the mass of a strange
star, so the discovery of a star with a minute radius, less than about 5 km,
or of a heavy planet of an enormous density, would
immediately confirm the existence of quark matter. An
 unusually high rotation rate could also reveal a quark star.

 A precise determination of the mass
is in principle possible through the measurement of the orbital
period in an accretion disk (as in the supermassive black hole candidate
NGC4258, where the velocity of water masers at various distances
from the central object has been measured: Miyoshi
et al., 1995). In LMXBs, it is currently not feasible to
 resolve spatially any ``blobs'' orbiting
in the disk; however, the mass of the compact
star can be determined from the maximum orbital frequency, $f_{orb,max}$,
 on the assumption that this maximum
is attained in the marginally stable orbit of general relativity
(Klu\'zniak et al. 1990).
The recently discovered kHz QPOs can in principle be used to determine
the orbital frequencies. In the simplest model the QPO frequencies
are the orbital frequencies and, on the assumption just stated,
the mass of the compact star in some sources
can be thus determined
(e.g. Kaaret et al. 1997, Zhang et al. 1998, Klu\'zniak 1998,
Bulik et al. 1999b).
Such considerations make important determining the 
range of possible orbital periods for various models of quark
stars (Stergioulas et al. 1999; Zdunik et al 2000a,b)
 and of neutron stars (Thampan et al. 1999).

Here, we summarize our studies of quark star masses
and frequencies (rotational and orbital), and  point out a new
complication: the marginally stable orbit exists
also in Newtonian gravity! If higher mass multipoles are present
in the distribution of matter, estimates of the central
mass from the maximum orbital frequency alone are unreliable.
As we show, the same value of $f_{orb,max}$ can correspond
to two stars differing in mass by a factor of a hundred, or more.
It is crucial to know the rotational period of a quark star, if its
mass is to be even approximately determined from the orbital frequency.

\begin{figure}
\centering
\leavevmode
%\epsfxsize=15cm
%\epsfysize=15cm
%\epsfbox{Mb01.eps}
\includegraphics[width=1.0\linewidth]{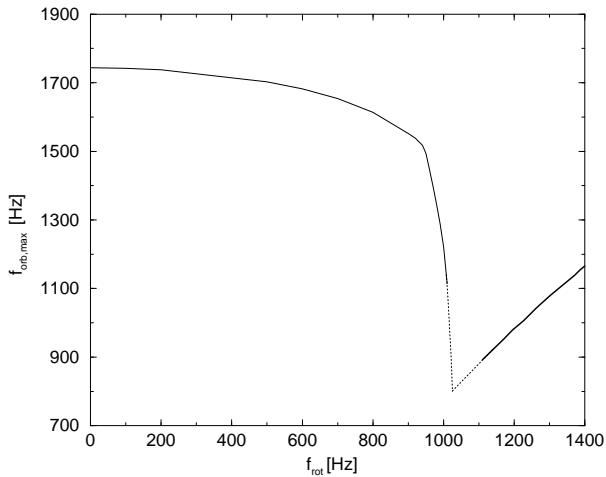}
\caption{Maximum orbital frequency for test particles in circular
motion in the equatorial plane
of a quark star with baryonic mass $M_b=0.01M_\odot$,
as a function of stellar spin rate (thin continuous line).
At low rotation rates, Keplerian orbits graze
the stellar surface, but for $f_{rot}>970\,$Hz
a gap separates stable circular orbits from the stellar
surface, and the maximum orbital frequency is attained in the marginally
stable orbit (ISCO).
Also shown are the ISCO frequencies for
maximally rotating quark stars
(thick continuous line, from Stergioulas et
al.~1999).
All calculations were performed with fully relativistic codes
for the MIT-bag model, with
$a=1/3, \rho_0=4.28\times 10^{14}\,{\rm g/cm}^3$, in eq. [1].
The dotted lines are extrapolations of these calculations
to lower orbital frequencies.
For maximally rotating models,
$f_{rot}$ is a monotonically increasing function of mass,
 all stars on the thick continous line have $M>1.3M_\odot$.
 Thus, the same value of $f_{orb,max}$ (e.g.,
1.1 kHz) can correspond  to  the ISCO around
a $0.01M_\odot$ star rotating at 1kHz,
or a $2M_\odot$ star rotating at about 1.3 kHz. 
}
\label{fig1}
\end{figure}
\section{Two models of quark matter and of strange stars} 

Traditionally, strange stars were modeled with an equation
of state based on the MIT-bag model of quark matter (Farhi and Jaffe 1984),
in which quark confinement is described by an energy term proportional
to the volume. Recently, Dey et al. (1998) proposed a model in which
quark interactions and confinement are described in a more self-consistent
manner. In fact, both models give a simple equation of state
$$P=a(\rho-\rho_0)c^2, \eqno (1)$$
 where $a$ is a constant of model-dependent
value (close to, but generally not equal to, $1/3$ for the MIT-bag model,
Zdunik 2000). It is the value, $\rho_0$, of density at zero pressure
which is crucial for various limiting properties of strange stars;
e.g., the maximum mass of static strange stars is
$M=2M_\odot(\rho_0/4\times10^{14})^{-1/2}$ in the MIT-bag model
(Witten 1984),
 and the scaling of mass with $\rho_0^{-1/2}$ is general.

Below, we describe our main results obtained for MIT-bag model stars.
For other models of quark stars the results are qualitatively similar,
However, in general
the Dey et al. (1998) model gives more compact stars, with lower
maximum mass and lower radii than the MIT-bag model stars. This
also translates to higher possible rotational and orbital frequencies
for the ``Dey stars'' (Gondek-Rosi\'nska et al. 2000a,b).

\section{Summary of results}

We were mostly interested in the value of maximum orbital frequency
of test particles (or fluid elements in an accretion disk) around such
stars. A very specific question was asked (Bulik et al. 1999a,b)
 whether the observed  maximum
frequency of kHz QPOs in 4U 1820-30, 1.07 kHz, is compatible with
strange star models---for static stars such a frequency is obtained
in the marginally stable orbit only for the relatively high mass of nearly
$2M_\odot$, according to the formula for the maximum orbital
frequency in Schwarzschild
metric, $f_{ms}=2198\, {\rm kHz}(M_\odot/M)$.
We found, that with the current physical constraints on $\rho_0$ (eq. [1])
masses of static strange stars as high as $2.4M_\odot$ cannot be ruled out
(Zdunik et al. 2000a).

To gauge the range of allowed orbital frequency, Stergioulas et al. (1999)
computed models of maximally rotating quark stars. The main results are,
that for these models the innermost (marginally) stable circular orbit (ISCO)
is always well above the stellar surface (even though rapid rotation leads
to a large increase of the equatorial radius), and that the ISCO frequencies
are usually much lower for rapidly rotating stars than for static ones, with
a value as low as 950 Hz for a $1.4M_\odot$ star rotating at the equatorial
mass shedding limit.
Only close to the maximum mass do ISCO frequencies in maximally
rotating quark stars become comparable to the static values.
 However, the ratio $T/W$ in these stars exceeds
the Newtonian limit for instability ($T/W > 0.1375$), so probably these
 extreme configurations
cannot exist. Detailed models of quark stars rotating at all possible rates
can be found in Zdunik et al. 2000b.

The gap between the ISCO and the stellar surface is always present for
quark stars more massive than $1.4$ to $1.6 M_\odot$ (depending on the model).
At somewhat lower masses the gap disappears for moderate rotation rates
(but the marginally stable orbit is present for both slowly and rapidly
rotating stars of this mass). It is only for the lowest mass stars,
that the gap disappears even at zero rotation rate. 

\section{The low mass limit}

The limit $M\rightarrow 0$ corresponds to $\rho \rightarrow\rho_0$ in eq. (1),
i.e., the star tends to a nearly uniform density configuration.
At the same time, the dimensionless angular momentum parameter
$j=cJ/(GM^2)\rightarrow\infty$, so an expansion in powers of angular
momentum (Klu\'zniak and Wagoner 1985, Shibata and Sasaki 1998)
is impossible; however, the limit is, in fact, Newtonian.
In Fig. 1 we show the maximum value of orbital frequency, as a function
of stellar rotation rate, for a low mass quark star (the curve is
nearly universal for all $M< 0.1M_\odot$). The computation was carried out
with a fully relativistic code, described in Gourgoulhon et al. (1999).

For all periods of rotation larger than about $1.03$ ms,
there are stable circular orbits just outside the surface. For non-rotating
stars, $f_{orb,max}$ is simply given by Kepler's law, and its value
decreases as the equatorial radius increases for rotating stars.
The cusp in the curve at about 970 Hz corresponds to the appearance
of a gap between the star and the ISCO. In this case,
this is a purely Newtonian effect related to the flattening of the star
at rapid rotation rates (see also Zdunik and Gourgoulhon 2000).

To demonstrate the existence of marginally stable orbits in Newtonian
theory, we present in Fig. 2 the angular momentum in circular orbit
around a uniform density disk, computed from the Newtonian gravitational
potential of this non-spherical mass distribution.

The conclusion is,
 that because of the appearance of the marginally stable orbit
in Newtonian gravity (when octupole and higher mass multipoles are
sufficiently large), the maximum orbital frequency can reach fairly low
values even for extremely low-mass stars. Conversely, $f=1.1\,$kHz
in the marginally stable orbit
need not imply that $M=2M_\odot$, we have exhibited a specific example
where the mass is $<0.01M_\odot$, instead. 

\begin{figure}
\centering
\leavevmode
\includegraphics[width=1.0\linewidth]{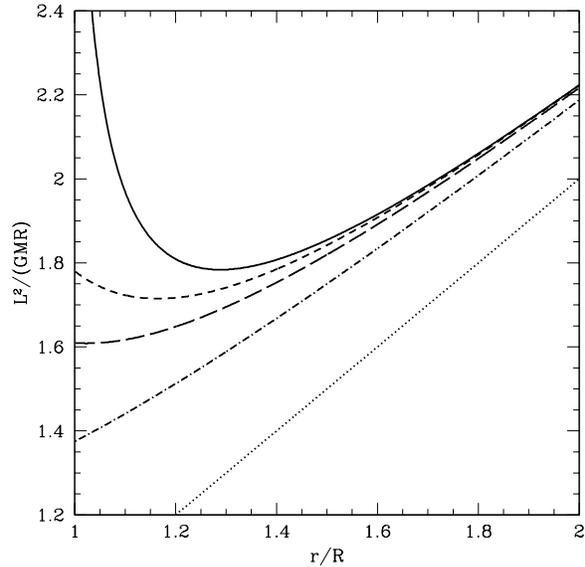}
\caption{Specific angular momentum squared, $L^2$, in Newtonian theory
of circular
orbits around a homogeneous disk of radius $R$ and mass $M$,
 as a function  of the orbital radius $r$ (continuous line).
Also shown are succesive approximations to this angular momentum,
when the multipole expansion of the Newtonian potential is truncated
at the monopole term (dotted line), and at the quadrupole, octupole,
16-pole terms (successively higher curves). Note that for this disk,
in purely Newtonian gravity, the marginally stable orbit---which
corresponds to the minimum of $L^2$---appears (just outside the disk)
 already in the octupole approximation, but
 actually occurs at  $r\approx 1.3R$, with the corresponding orbital frequency,
$2\pi f_{orb,max}=L/r^2$, 
about 20\% higher than for a circular orbit (of the same radius)
around a spherical mass distribution of mass $M$. On the other hand,
the same $f_{orb,max}$ is only 0.8 of the value of the orbital frequency
at $r=R$ around a spherical mass $M$.
}
\label{fig2}
\end{figure}
%
%\section*{ACKNOWLEDGMENTS}
This work has been funded by the KBN grants 2P03D01816, 2P03D00418.
%The numerical calculations have been performed on
%computers purchased thanks to a special grant from the SPM and SDU
%departments of CNRS.
%
\section*{REFERENCES}

Alcock, C., Farhi, E. and Olinto, A., 1986, ApJ 310, 261

Alpar, M.A., 1987, Phys. Rev. Lett. 58, 2152

Bodmer, A.R., 1971, Phys. Rev. 4, 1601

Brecher, K. and Caporaso, G., 1976, Nature 259, 377

Bulik, T., Gondek-Rosi{\'n}ska, D., and Klu\'zniak, W., 1999a, Astrophys.
Lett. Commun. 38, 205

Bulik, T., Gondek-Rosi{\'n}ska, D., and Klu\'zniak, W., 1999b, A\&A, 344, L71

Dey, M., {Bombaci}, I., {Dey}, J., {Ray}, S., and {Samanta}, B.~C., 1998,
Physics Letters B, {438}, 123

Farhi, E. and Jaffe, R.L., 1984, Phys. Rev. D 30, 2379

Gondek-Rosi{\'n}ska, D., Bulik T., Zdunik J. L., Gourgoulhon E., Ray S.,
 Dey J., Dey M., 2000a, A\&A, in press

Gondek-Rosi{\'n}ska, D., Bulik T., Klu\'zniak, W., Zdunik J. L.,
 Gourgoulhon E., 2000b, these proceedings

Gourgoulhon, E., {Haensel}, P., {Livine}, R., {Paluch}, E., {Bonazzola}, S.,
  and {Marck}, J.~A., 1999, A\&A, {349}, 851

Haensel, P., Zdunik, J.L. and Schaefer, R., 1986, A\&A 160, 121

Kaaret, P., Ford, E. C., Chen, K. 1997, ApJ, 480, 127

Klu\'zniak, W., 1998, ApJ, 509, L37

Klu\'zniak, W., and Wagoner, R. V., 1985, ApJ, 297, 548

Klu\'zniak, W., Michelson, P. and Wagoner, R. V., 1990, ApJ, 358, 538

Miyoshi, M. et al., 1995, Nature 373, 127

Shibata, M. and Sasaki, M. 1998

Stergioulas, N., Klu{\'z}niak, W., and Bulik, T., 1999, A\&A, 352, L116

Thampan, A.V., Bhattacharya, D. and Datta, B., 1999, MNRAS 302, L69

Witten, E. 1984, Phys. Rev. 30, 272

Zdunik, J. L., 2000, A\&A, 359, 311

Zdunik, J.\ L., Bulik, T., Klu{\'z}niak, W., Haensel, P. and
Gondek-Rosi{\'n}ska, D.\ 2000a, A\&A, 359, 143 

Zdunik, J. L., {Haensel}, P., {Gondek-Rosi{\'n}ska}, D. and {Gourgoulhon}, E.,
 2000b, {A\&A} {356}, 612 

Zdunik, J. L. and  {Gourgoulhon}, E., 2000, astro-ph/0011028

Zhang, W., Smale, A.P., Strohmayer, T. E. and
 Swank, J. H. 1998, ApJLet 500, L171

\end{document}